\documentclass[twoside]{article}

\usepackage[hmarginratio=1:1,top=20mm,left=20mm,right=20mm,columnsep=20pt]{geometry} 
\usepackage{multicol} 
\usepackage[hang, small,labelfont=bf,up,textfont=it,up]{caption} 
\usepackage{booktabs} 
\usepackage{float} 
\usepackage[pdftex]{hyperref}
\hypersetup{
pdftitle={Money Walks: A Human-Centric Study on the Economics of Personal Mobile Data},
pdfauthor={Jacopo Staiano, Nuria Oliver, Bruno Lepri, Rodrigo de Oliveira, Nicu Sebe},
pdfkeywords={Privacy; Economics; Personal Mobile Data; Auction; Mobile Computing; Living Lab; Monetization},
bookmarksnumbered,
pdfstartview={FitH},
colorlinks
}
\usepackage{graphicx}

\usepackage{paralist} 

\usepackage{abstract} 

\usepackage{titlesec} 
\renewcommand\thesection{\Roman{section}} 
\renewcommand\thesubsection{\Alph{subsection}}
\titleformat{\section}[block]{\bf \scshape\centering}{\thesection.}{1em}{} 
\titleformat{\subsection}[block]{\bf \centering}{\thesubsection.}{1em}{} 
\titleformat{\subsubsection}[block]{\it}{\thesubsubsection.}{1em}{} 

\usepackage{multirow}
\usepackage{algpseudocode}
\usepackage{textcomp} 

\usepackage{times}
\usepackage{url}
\usepackage{caption} 
\usepackage{booktabs} 
\usepackage{float} 

\usepackage[official]{eurosym}
\newcommand*\justify{%
  \fontdimen2\font=0.4em
  \fontdimen3\font=0.2em
  \fontdimen4\font=0.1em
  \fontdimen7\font=0.1em
  \hyphenchar\font=`\-
}

\makeatletter
\def\blfootnote{\xdef\@thefnmark{}\@footnotetext}
\makeatother

\title{\vspace{-15mm}\fontsize{22pt}{10pt}\selectfont\textbf{Money Walks: A Human-Centric Study on the Economics of Personal Mobile Data}} 

\author{Jacopo Staiano$^\dag$, Nuria Oliver$^\ddag$, Bruno Lepri$^\S$,\\
Rodrigo de Oliveira$^\ddag$, Michele Caraviello$^\neg$, Nicu Sebe$^\dag$ \\
\textit{\small $^\dag$DISI, University of Trento (Italy)}\\\textit{\small $^\ddag$Telefonica Research, Barcellona (Spain)}\\\textit{\small $^\S$Fondazione Bruno Kessler, Trento
(Italy)}\\
\textit{\small $^\neg$Telecom Italia, Trento (Italy)}\\
{\small The work of Jacopo Staiano has been supported by and performed at Telefonica Research.}\\
{\small Rodrigo de Oliveira is currently afﬁliated with Google Inc., USA. All research conducted while he was at Telefonica Research, Barcelona.}}
\date{}

\begin{document}

\maketitle 



\begin{abstract}
In the context of a myriad of mobile apps which collect personally identifiable information (PII) and a prospective market place of personal data, we investigate a user-centric 
monetary valuation of mobile PII. During a 6-week long user study in a living lab deployment with 60 participants, we collected their daily valuations of 4 categories of mobile 
PII 
(communication, \emph{e.g.} phonecalls made/received, applications, \emph{e.g.} time spent on different apps, location and media, \emph{e.g.} photos taken) 
at three levels of 
complexity 
(individual data points, aggregated statistics and processed, \emph{i.e.} meaningful interpretations of the data). 
In order to obtain honest valuations, we employ a
reverse second price auction mechanism. Our findings show that the most sensitive and valued category of personal information is location. We report statistically significant 
associations between actual mobile usage, personal dispositions, and bidding behavior. 
Finally, we outline key implications for the design of mobile services and future markets of personal data. 

\end{abstract}


\begin{multicols}{2} 

\section{Introduction}
The number of mobile phones actively in use worldwide today is about 5 billion, with millions of new subscribers every 
day\footnote{\url{http://www.ericsson.com/ericsson-mobility-report}\\}. Mobile phones allow for unobtrusive and cost-effective access to previously inaccessible sources of 
behavioral data such as location, communications (calls and text messages), photos, videos, apps and Internet access \cite{LaneEtAl2010}. Hence, 
a result of the ever-increasing adoption of these devices is the availability of large amounts of \emph{personal data} related to habits, routines, social interactions and 
interests
\cite{LaneEtAl2010,MadanEtAl2012}. 

However, the ubiquitous collection of personal data raises unprecedented privacy challenges. Users typically have to make decisions concerning the disclosure of their 
personal information on the basis of a difficult tradeoff between data protection and the advantages stemming from data sharing. Perhaps more importantly, people are typically not
involved in the life-cycle of their own personal data -- as it is collected by websites and mobile phone apps, which results in a lack of understanding of who uses their data and 
for what. 

Several researchers have proposed and investigated new user-centric models for personal 
data management, which enable individuals to have more control of their own data's life-cycle \cite{Pentland2012}. 
To this end, researchers and companies are developing repositories which implement medium-grained access control to different kinds of 
personally identifiable information (PII), such as \emph{e.g.} passwords, social security 
numbers and health info \cite{WantEtAl2002}, and more recently location \cite{yves2012trusted, HongLanday2004, MunEtAl2010} and personal data collected online 
by means of smartphones or wearable devices \cite{yves2012trusted, VanKleekEtAl2012}.

Previous work has introduced the concept of \emph{personal data markets} in which individuals sell their own personal data to entities interested in buying it 
\cite{AdarHuberman2001}. Buyers are likely to be companies and researchers, while sellers are individuals who receive compensation for 
sharing their own data. Riederer \emph{et al.}~\cite{Riederer2011} have recently proposed a mechanism called \emph{transactional} privacy, 
devised to maximize both the user's control of their own PII and the utility of a data-driven market.

In the context of prospective personal data markets 
that offer increased transparency and control, it is of great importance to understand the value that users put to their own PII. 
Recently, Carrascal \emph{et al.}~\cite{carrascal2013} used a refined 
Experience Sampling Method (rESM) \cite{cherubini2009} and a reverse second price auction to assess the monetary value that people assign to their PII shared online via websites 
-- \emph{e.g.} keywords used in a search engine, photos shared in a social network, etc. However, the authors focus only on web-browsing behaviors without taking into account 
behaviors and personal information that can be captured by mobile phones. 

Taking Carrascal \emph{et al.}~\cite{carrascal2013} as an inspiration, in this paper 
we investigate the monetary value that people assign to different kinds of PII as collected by their mobile phone, including location and communication information.
We carried out a comprehensive 6-week long study in a living lab environment with 60 participants
and adopted a Day Reconstruction Method \cite{kahnemanEtAl2004} along with a reverse second price auction mechanism in order to poll and collect honest monetary valuations.

The main contributions of this paper are:

\begin{enumerate}
\item 
Quantitative valuations of mobile PII as collected by a 6-week long study conducted in the wild;
\item 
Qualitative feedback on the valuations provided by each participant as gathered by an End of Study (EoS) survey; 
\item 
A segmentation of PII valuations and findings based on 4 categories of mobile PII (communications, location, media and apps), 3 levels of complexity (individual, processed, 
aggregated), and one level of temporal granularity (daily);
\item 
A set of key insights about people's sensitivities and valuations of mobile PII and implications for the design of mobile services that leverage mobile PII. 
\end{enumerate}

\section{Related Work} 
\label{sec:rel_works}

In recent years, researchers have analyzed the factors that can influence a person's disclosure behavior and economic valuation of personal 
information. Demographic characteristics, such as gender and age, have been found to affect disclosure attitudes and behavior. Several studies have identified gender differences 
concerning 
privacy concerns and consequent information disclosure behaviors: for example, women are generally more protective of their online privacy \cite{FogelNehmad2009, HoyMilne2010}. 
Age 
also plays a role in information disclosure behaviors: in 
a study on Facebook usage, Christofides \emph{et al.} \cite{ChristofidesEtAl2011} found that adolescents disclose more information.

Prior work has also emphasized the role of an individual's stable psychological attributes - \emph{e.g.} personality traits - to explain information disclosure behavior. 
Korzaan \emph{et al.} \cite{KorzaanEtAl2009} explored the role of the Big5 personality traits \cite{CostaMcCrae2005} and found that Agreeableness --
defined as being sympathetic, straightforward and selfless, has a significant influence on individual concerns for information privacy. 
Junglas \emph{et al.} \cite{JunglasEtAl2008} and Amichai-Hamburger and Vinitzky \cite{Amichai2010} also used the Big5 personality traits and 
found that Agreeableness, Conscientiousness, and Openness affect a person's concerns for privacy. 
However, other studies targeting the influence of personality traits did not find significant correlations 
\cite{SchrammelEtAl2006}. More recently, Quercia \emph{et al.} \cite{QuerciaEtAl2012} found weak correlations among Openness to Experience and, to a lesser extent, Extraversion 
and 
the 
disclosure attitudes on Facebook. In 2010, Lo \cite{Lo2010} suggested that Locus of Control \cite{Rotter1967} could affect 
an individual's perception of risk when disclosing personal information: internals are more likely than externals to feel that they can control the risk of becoming privacy 
victims, hence they are more willing to disclose their personal information \cite{YoungQuanHaase2009}. 

Individual differences are also found when providing economic valuations of 
personal data \cite{AcquistiEtAl2009, carrascal2013}. 
For instance, some individuals may not be concerned about privacy and would 
allow access to their data in exchange for a few cents, whereas others may only consent if well paid. Recently, Aperjis and Huberman \cite{AperjisHuberman2012} proposed 
to introduce a realistic market for personal data that pays individuals for their data while taking into account their own privacy and risk attitudes. 

Previous research has shown that disclosure \cite{KnijnenburgEtAl2013} and valuation \cite{DanezisEtAl2005,HubermanEtAl2005} depend on the kind of information to be 
released. Huberman \emph{et al.} \cite{HubermanEtAl2005} reported that the valuation of some types of personal information, such as the subject's weight and the subject's age 
depends on 
the desirability of these types of information in a social context. Some empirical studies have attempted to quantify subjective privacy valuations of personal information in 
different contexts,  
such as personal information revealed online \cite{HannEtAl2007}, access to location data \cite{CvrcekEtAl2006}, or removal from marketers' call lists \cite{VarianEtAl2005}. These 
studies can be classified into two groups. The first and larger group includes studies that explicitly or implicitly measure the amount 
of money or benefit that a person considers to be  
enough to share her/his personal data, namely their \emph{willingness to accept} (WTA) giving away his/her own data (see for 
example \cite{CvrcekEtAl2006, HuiEtAl2007}). The second and smaller group 
includes studies about tangible prices or intangible costs consumers are \emph{willing to pay} (WTP) to protect their privacy (see for example, 
\cite{AcquistiGrossklags2005, TsaiEtAl2011}). In our paper, we do not deal with WTA vs WTP, but we focus on WTA for PII captured by mobile phones (communications, apps and media 
usage, locations).

A growing body of studies in the fields of ubiquitous and pervasive computing and human-computer interaction focuses on location sharing behavior 
and has highlighted the role played by the recipient of sharing (who can access the information), the purpose, the context, how the information is going to be used 
\cite{BarkhuusEtAl2008, ConsolvoEtAl2005, LindqvistEtAl2011, TochEtAl2013, WieseEtAl2011} and the level of granularity of the information shared \cite{LinEtAL2010}.
Finally, studies have suggested the importance of analyzing people's actual behavior rather than attitudes expressed through questionnaires because often the actual behavior of 
people deviates from what they state \cite{JensenEtAl2005}. 

Building upon previous work, in this paper we investigate 
the monetary value that people assign to different kinds of PII as collected by their mobile phone, including location and communication patterns. 
In particular, we carry out a comprehensive 6-week long  
study in a living lab environment with 60 participants
and adopt a Day Reconstruction Method \cite{kahnemanEtAl2004} and a reverse second price auction mechanism in order to poll and collect honest monetary valuations from our sample. 

\section{Methodology}
\label{sec:methodology}
Next, we describe the methodology followed during our 6-week study.

\subsection{The Living Laboratory}
The Living Laboratory where we carried out our study was launched in November of 2012 and it is a joint effort between industrial and academic research institutions. 
It consists of a group of more than 100 volunteers who carry an instrumented smartphone in exchange for a monthly credit bonus of voice, SMS and data access. 
The sensing system installed on the smartphones is based on the FunF\footnote{\url{http://funf.org}} framework \cite{aharony2011social} and logs communication events, location, 
apps usage and 
photos shot. In addition, 
the members of the living lab participate in user-studies carried out by researchers.  
The goals of this living lab are to foster research on real-life behavioral analysis obtained by means of mobile devices, 
and to deploy and test prototype applications in a real-life scenario. One of the most important features of such a lab is its ecological validity, given that the participants' 
behaviors and attitudes are sensed in the real world, as people live their everyday life, and not under artificial laboratory conditions.

All volunteers were recruited within the target group of young 
families with children, using a snowball sampling approach where existing study subjects recruit future subjects from among their acquaintances \cite{Goodman1961}. 
Upon agreeing to the terms of participation, the volunteers granted researchers legal access to their behavioral data as it is collected by their smartphones. Volunteers retain 
full rights over their personal data such that they can order deletion 
of personal information from the secure storage servers. Moreover, participants have the choice  to participate or not in a given study.
Upon joining the living lab, each participant fills out an initial questionnaire which collects their demographics, individual traits and dispositions (\emph{e.g.} Big Five 
personality traits, trust disposition, Locus of Control, etc.) information.

\subsection{Participants}
A total of 60 volunteers from the living lab chose to participate in our mobile personal data monetization study. 
Participants' age ranged from 28 to 44 years old ($\mu=38$, $\sigma=3.4$). They held a variety of occupations and 
education levels, ranging from high school diplomas to PhD degrees. All were savvy Android users who had used the smartphones provided by the living lab  
since November 2012. Regarding their socio-economic status, the average personal net income amounted to \euro{21169} per year ($\sigma=5955$); while the average family net 
income amounted to \euro{ 36915} per year ($\sigma=10961$). All participants lived in Italy and the vast majority were of Italian nationality.

\subsection{Procedure}
Our study ran for six weeks from October 28th, 2013 to December 11th, 2013. At the beginning of the study, participants were explained that the study consisted of three phases:
\begin{enumerate}
 \item An initial questionnaire, which focused on their general perception of privacy and personal data;
 \item A daily data collection phase that lasted 6 weeks where participants answered daily surveys to valuate their mobile personal data; 
 \item A final survey that aimed to clarify the results obtained and to collect qualitative feedback from participants.
\end{enumerate}

\paragraph{Daily Surveys}
Ad-hoc \texttt{java} code was developed and scheduled to run on a secure server each night in order to automatically generate personalized daily surveys for each participant. The 
survey questions were generated based on the mobile data collected during the previous day. 
Everyday, at 12PM, participants received an SMS reminding them to fill out their survey via a personalized URL (through a unique hash). 

In order to test the live system and identify bugs, we ran a pilot for 10 days with a small set of volunteers who were not participants in the study. In addition, we allocated a 
\emph{training} week prior to starting the actual study so participants would get accustomed to the survey/auction scheme.

\section{Collected Data}
Next we describe the data that we collected during the study.

\subsection{Mobile Personal Data}
We collected 4 categories of mobile personal data: 
(1) \emph{communications}, in the form of calls made/received; 
(2) \emph{locations}, collected by the device GPS sensor every $\sim5$ minutes; (3) \emph{running applications}, sampled every 25 minutes; and 
(4) \emph{media}, \emph{i.e.} number and timestamp of pictures taken and obtained by monitoring the device file system. 
The sampling rates for the different categories of data were empirically determined in order to have good resolution without significantly impacting the device's battery life.

Moving from finer to coarser granularity, we probed participants about the following three levels of 
complexity for each category of data:
(1) \emph{individual}, encompassing individual data points (\emph{e.g.} a call made/received, a picture taken, a specific GPS location); 
(2) \emph{processed}, depicting higher level information derived from the sensed events (\emph{e.g.} a given application has been running 
for N minutes, total distance traveled); and (3) \emph{aggregated}, portraying cumulative event information (\emph{e.g.} number of places visited, number of calls 
made/received).

For each data category and level of complexity, participants were asked to fill out daily surveys that asked them about data from the previous day for each category and for a 
specific level of complexity (up to 4 questions per day). For each question in the surveys, participants always had the option to opt-out and not sell that particular piece of 
information. 

Next, we describe in detail the 4 categories and the 3 levels of complexity of mobile personal data that we 
collected in this study, which are summarized in Table~\ref{tab:features}. 

\begin{table*}
\centering
 \begin{tabular}{l||r|r|r}
  \textbf{Category}&\textbf{Individual}&\textbf{Processed}&\textbf{Aggregated}\\
\hline
  Communications&A call event [*]&Total duration of calls&\# of calls or diversity\\
  Location&A place visited [*]&Total distance covered&\# of places visited\\
  Running Apps&App \emph{X} running [*]&App \emph{X} running for \emph{N} minutes in the [**]&\# of apps running\\
  Media&A picture shot [*]&Pictures shot in the [**]&\# of pictures shot\\
 \end{tabular}
\caption{\small Categories of personal data probed in the surveys. Include [*: at time hh:mm; **: night (12AM-6AM), morning (6AM-12PM), afternoon (12PM-18PM), evening 
(18PM-12AM)]. All questions referred to data collected the previous day.} 
\label{tab:features}
\end{table*}

\subsubsection[]{Communications}
\emph{Individual} communication data was restricted to voice calls made/received; missed calls were discarded. 
The \emph{processed} communication variable referred to the total duration of calls in the previous day, 
resulting in questions such as 
\texttt{"Yesterday, you spoke on the phone for a total of 52 minutes"}.

With respect to \emph{aggregated} communications data, we alternated between two different aggregated variables on a weekly basis: 
on even weeks subjects were asked to monetize information about the total number of calls made/received during the previous day,
while on odd weeks they were asked about call diversity, \emph{i.e.} the number of different people that they talked to on the phone during the previous day.
Examples of questions related to aggregate communications are 
\texttt{"Yesterday, you made/received 8 phone calls"},  
or 
\texttt{"Yesterday, you spoke on the phone with 3 different persons"}.

\subsubsection[]{Location}
\emph{Individual} location referred to a specific place visited by the participant in the previous day. Semantic information associated to GPS locations was derived via reverse 
geo-coding using Yahoo Query Language. For individual locations, details on street, neighborhood and town were included in the question. 
For example, \texttt{"Yesterday, at 23:56 you were in Via Degli Orbi 4, Trento"}.
The \emph{processed} location variable referred to the total distance traveled in the previous day,
resulting in questions such as \texttt{"Yesterday you covered a total distance of 13km"}.

Finally, location data was spatially clustered over the reference time-range using a threshold of 100 meters to generate the \emph{aggregated} location question (\emph{e.g.} 
\texttt{"Yesterday you have been in 23 different places"}).

\subsubsection[]{Running Applications}
With respect to running apps, the \emph{individual} variable included 
the timestamp and the name of the app running in the foreground.
\emph{Processed} app information referred to the total number of minutes that a particular app was running over a specific time in the previous day,
whereas
\emph{aggregated} app variables referred to the total number of different apps that the participant ran the previous day.

Examples of questions on app-related information for each level of complexity are \texttt{\justify "Yesterday, at 10:23 you were using the Firefox Browser application"},
\texttt{\justify "Yesterday night, the Google Talk application run on your device for 82 minutes"}, and \texttt{"Yesterday 9 applications were running on your device"}, 
respectively. 

\subsubsection[]{Media}
\emph{Individual} media asked participants about the fact that they shot a photo at a specific time 
(\texttt{"Yesterday, at 14:23, you shot one picture"}). 
For legal privacy reasons, the questions referring to individual media data could not include the actual picture they referred to.
\emph{Processed} media 
probed participants about their photo-taking activity during specific times of the day (\emph{e.g.} \texttt{"Yesterday morning you took 4 pictures"}).
Finally, the \emph{aggregated} media variable referred the total number of pictures shot the previous day
(\emph{e.g.} \texttt{"Yesterday you took 9 pictures"}). 

\subsection{Individual Traits Data}
As previously mentioned, upon joining the lab each participant filled out 4 questionnaires to collect information about their personality, locus of control, dispositional trust 
and 
self-disclosure behaviors. 

The Big Five personality traits were measured by means of the BFMS \cite{PeruginiDiBlas2002} questionnaire, which is validated for the Italian language and covers the traditional 
dimensions of Extraversion, Neuroticism, Agreeableness, Conscientiousness, and Openness \cite{CostaMcCrae2005}. 
Participants also provided information about their \emph{Locus of Control} (LoC) \cite{Rotter1965}, a psychological construct measuring whether causal attribution for subject 
behavior or beliefs is made to oneself or to external events and 
circumstances. The LoC measures whether the outcomes of a set of beliefs are dependent upon what the subject does (internal orientation) or upon events outside of her/his control 
(external orientation). LoC was measured by the Italian version of Craig's Locus of Control scale \cite{FarmaCortivonis2000}. 

Moreover, we collected information about the participants' \emph{dispositional trust}. Rotter \cite{Rotter1967} was among the first to discuss trust as a form of personality 
trait, 
defining interpersonal trust as a generalized expectancy that the words or promises of others can be relied on. In our study, we resort to Mayer and Davis's Trust Propensity Scale 
\cite{MayerDavis1999}. 

Finally, we targeted the \emph{self-disclosure} attitudes of our subjects. Self-disclosure has been defined as any message about the self that an individual communicates to 
another 
one \cite{Cozby1973, WhelessGrotz1976}. We used Wheeless' scale \cite{WhelessGrotz1976} measuring five dimensions of \
emph{self disclosure}, namely (i) amount of disclosure, (ii) positive-negative nature of disclosure, (iii) consciously intended disclosure, (iv) honesty and accuracy of 
disclosure, 
and (v) general depth or intimacy of disclosure. Wheeless' scale has been utilized to measure self-disclosure in online communication and in interpersonal relationships 
\cite{WhelessGrotz1976}.

\subsection{Auctions of mobile PII}
The personalized daily survey asked each participant to place a bid to sell one piece of their mobile personal information for each of the four categories of study 
(communications, 
location, apps and media), for a specific level of complexity (individual, processed, or aggregated) and for the previous day. The winner of each auction won 
the monetary value associated with that auction. In exchange, (s)he sold 
that particular piece of information to the Living Lab which could 
use it for whatever purpose it wanted.

In order to ensure a balanced sample, surveys were generated by rotating the different levels of complexity described above, such that each day participants placed bids in up to 4 
auctions: one for each category of personal information and for a particular level of complexity (individual, aggregated or processed). Note that in the case a participant did not 
generate any data for a particular category, s(he) was still asked to 
provide a valuation to the fact that there was no data in that category, \emph{e.g.} 
\texttt{"Yesterday you did not make any phone call"}. 

The participants' bids entered a reverse second-price auction strategy, \emph{i.e.}, the winner was the participant(s) who bid the lowest, and the prize was the second lowest bid. 
The choice of this auction mechanism was due to the following reasons: (1) the mechanism is truth telling given that the best strategy for the auction participants is to be honest 
about their valuation \cite{McAfeeMcMillan1987}, (2) it is easy to explain and understand, and (3) it has successfully been used before to evaluate location information in 
\cite{DanezisEtAl2005} and Web-browsing information in \cite{carrascal2013}. 
\begin{table*}[!ht]
 \centering

\resizebox{\linewidth}{!}{
 \begin{tabular}{p{14cm}|c|c}
 \textbf{Question}&\textbf{mean}&\textbf{st\_dev}\\
 \hline
 \textbf{Q1.} I am concerned about the protection of the data collected by my smartphone &4.7&1.6\\
 \hline
 \textbf{Q2.} I trust the applications I install and run on my smartphone wrt how they use my data&3.7&1.5\\
 \hline
 \textbf{Q3.} I trust telco providers with respect to how they use my data&3.4&1.4\\
 \hline
 \textbf{Q4.} I always read the privacy terms and conditions for the applications I use&2.7&1.6\\
 \hline
 \textbf{Q5.} I know the legislation on mobile communication data protection&2.5&1.5\\
 \hline 
 \end{tabular}
}
\caption{\small Questions asked in the Initial questionnaire, and responses statistics. The 7-point likert scale used goes from \emph{1-Totally Disagree} to \emph{7-Totally 
Agree}.}
\label{tab:initsurvey}
\end{table*}
Interventions, \emph{i.e.} individual communications of auction outcomes to participants, took the form of e-mails sent every Thursday.

In order to evaluate possible effects of winning frequency on bidding behavior, we employed two different auction strategies for the first and second halves of the study. During 
the first 3 weeks (phase 1), we carried out weekly auctions on Wednesday, taking into account all bids that had been entered during the previous 7 days for each category. 
Therefore, in this phase, 12 weekly auctions took place with the daily bids for each category and level of complexity (4 categories x 3 levels of complexity).
During the last 3 weeks of the study (phase 2), we switched to daily auctions; furthermore, the sample of bidding 
participants was split into 3 random subsets in order to increase their chances of winning, resulting in a total of 12 auctions per day. 

Email interventions were always on Thursdays and therefore this change was transparent to participants. Interventions were sent to all participants, whether they had won auctions 
or not. In the case of winners, the intervention email included the specific piece of information that the participant had sold, the corresponding winning bid, and the amount won. 
In the case of losers, the intervention email simply communicated the participant that s(he) did not win any of their auctions. All emails were kept neutral for both winners and 
losers. 

In total, 596 auctions were run during the entire study 
(36 in the first three weeks, 560 afterwards).
\begin{table*}[!ht]
 \centering
\resizebox{\linewidth}{!}{
 \begin{tabular}{p{15cm}|c}
 \textbf{Question}&\textbf{Type}\\
 \hline
\textbf{Q1.} This \{map$\|$chart\} shows the information about \{locations$\|$communications$\|$apps$\|$media\} we collected during this study. 
 What is the minimum amount of money you would accept to sell it in anonymized/aggregated form?&numeric\\
 \hline
\textbf{Q2.} On day \{dd/MM\} you assigned a value of \{\emph{min-bid per category}\} to the information [\{\emph{least valued info per category}\}]. This was your minimum bid. 
Why?&multi-choice*\\
 \hline
\textbf{Q3.} On day \{dd/MM\} you assigned a value of \{\emph{max-bid per category}\} to the information [\{\emph{most valued info per category}\}]. This was your maximum bid. 
Why?&multi-choice*\\
\hline
\textbf{Q4.} Imagine there was a market in which you could sell your personal information (e.g. information about people you called, places you've been, applications you've used, 
songs you've 
listened to, etc.). Who would you trust to handle your information? Please, order the following entities from most to least trusted.&rank**\\
 \hline
\textbf{Q5.} The category \{locations$\|$communications$\|$apps$\|$media\} is the one that you refused to sell the most (\{\emph{percentage of opt-outs}\}). Why?&free-text\\
 \hline
 \end{tabular}
}
\caption{\small Questions asked in the EoS questionnaire. 
*included: \emph{Fair value}, \emph{Test/Mistake}, \emph{Other (free text)}. For minimum-bid related 
questions additional options were \emph{To win the auction}, \emph{Info not important}; conversely, for maximum-bid related questions, the additional option was \emph{To prevent 
selling}.
**entities to be ranked included: \emph{banks}, \emph{government}, \emph{insurance companies}, \emph{telcos}, \emph{yourself}. 
}
\label{tab:finalsurvey}
\end{table*}
\subsection{Pre- and Post- Study Questionnaires}
As previously explained, at the beginning and at the end of the data collection participants were required to fill out initial and end-of-study (EoS) questionnaires. 
The initial questionnaire consisted of 5 questions (see Table~\ref{tab:initsurvey}) 
and was used to gather information about the participants' perception of privacy issues related to mobile personal data.
From the responses provided to this survey, we notice that participants are concerned about mobile PII protection (Q1) but do not tend to read the Terms of Service (Q4) nor are 
aware of current legislation on data protection (Q5). Moreover, they do not seem to trust how
neither application providers (Q2) nor telecom operators (Q3) use their data. 

\begin{figure}[H]
\centering
\includegraphics[width=\columnwidth]{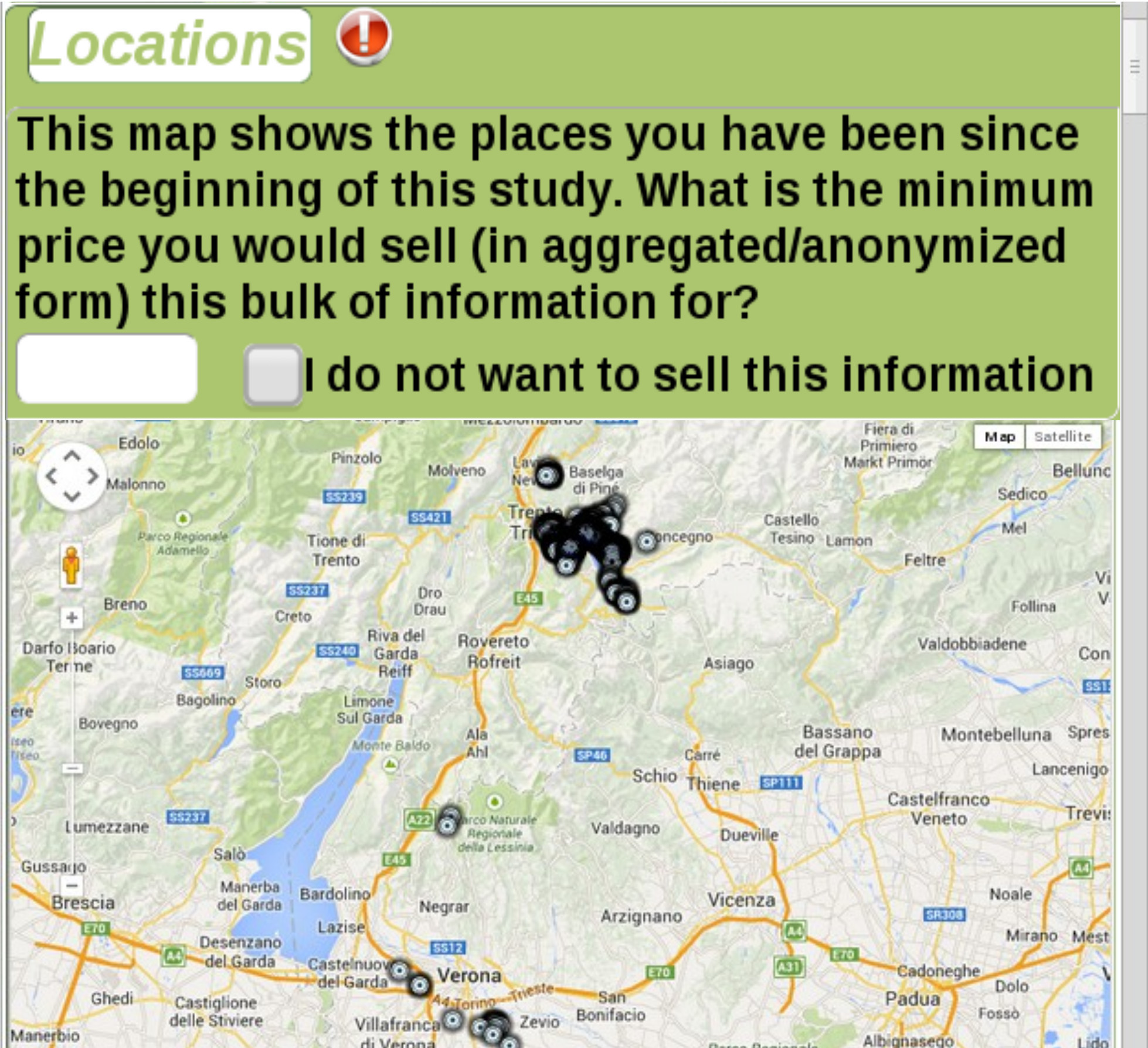}
\caption{Location-specific bulk information question in the EoS survey.} 
\label{fig:bulk-screen}
\end{figure}

The EoS survey was designed to gather additional quantitative and qualitative information from our participants after the data collection was complete. In particular, we asked 
participants to put a value (under the same auction game constraints) on category-specific \emph{bulk information} -- \emph{i.e.} all the data gathered in the study for each 
category. For instance, in the case of location information, a visualization of a participant's mobility data collected over the 6-weeks period was shown in the 
Web questionnaire (as depicted in Figure~\ref{fig:bulk-screen}) 
and the participant was asked to assign it a monetary value. 
Furthermore, for each category, we asked participants about the minimum/maximum valuations given during the study, in order to understand the reasons why they gave these 
valuations. Table~\ref{tab:finalsurvey} contains all the questions of the EoS survey.

The EoS questionnaire was administered through a slightly modified version of the same Web application used for the daily surveys. The main difference are the visualizations of 
the 
collected data.

\section{Data Statistics}
The data used throughout this paper was collected from October 28th and December 11th 2013, inclusive. 
Data was not collected for the first 3 days of November, due to the All Saints festivities in Italy; hence, our data-set encompasses 43 days. 

A total of 2838 daily surveys 
were administered during this period. Statistics on bidding data and participation follow.
\begin{figure*}[!ht]
\centering
\includegraphics[width=.7\linewidth]{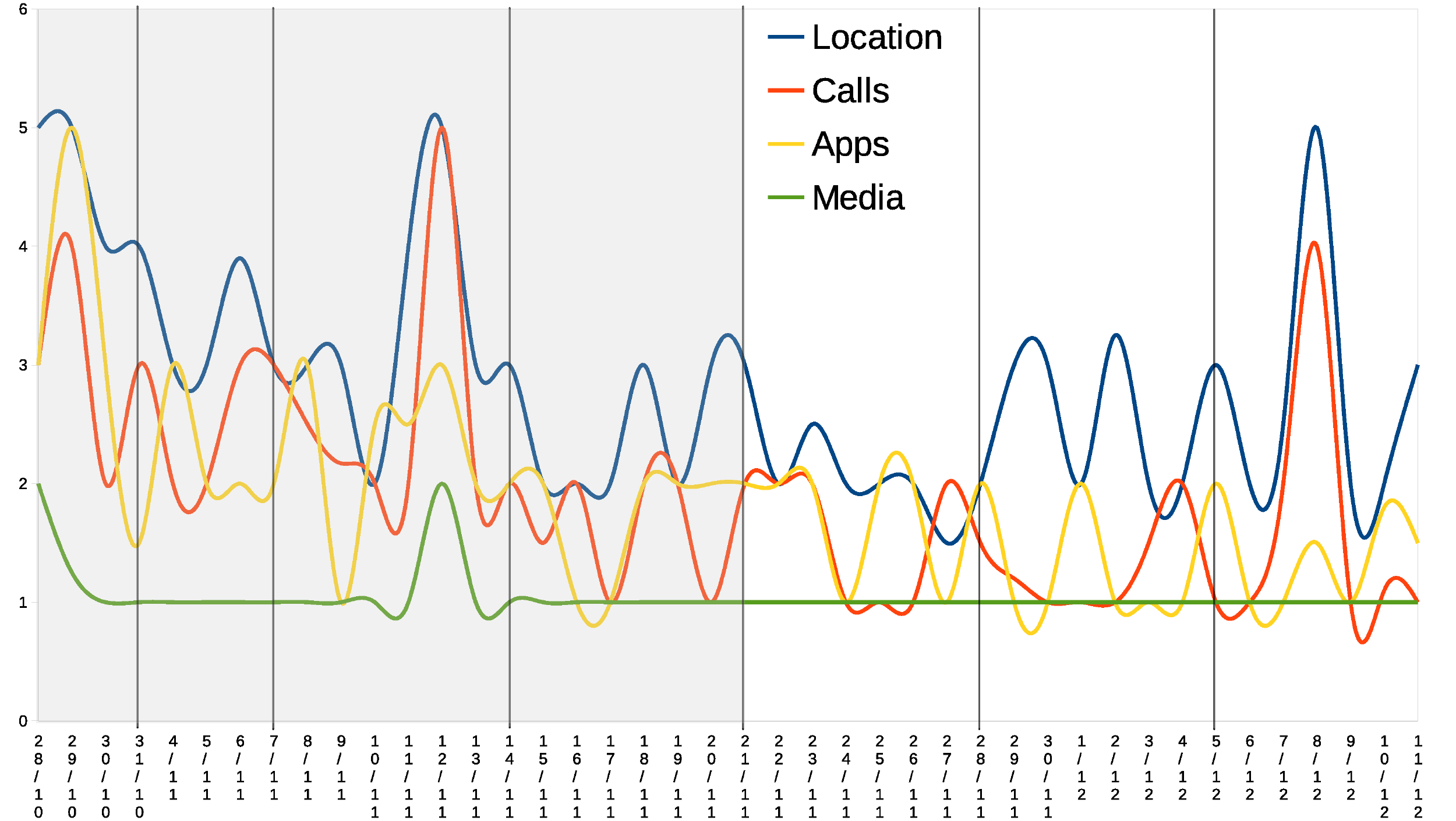}
\caption{Daily median bid values (\euro) per category. Vertical lines indicate interventions. Shaded area indicates phase 1.}
\label{fig:medianbids}
\end{figure*}
\subsection{Bids}
Table \ref{tab:bid-values}
summarizes the bidding values for each personal data category and level of complexity. 
Figure~\ref{fig:medianbids} depicts median bid values each day for
each category and level of complexity\footnote{Note how the spatial gap between the first two interventions is smaller than between the rest of interventions because of the lack 
of 
data
during 3 days in November.}.

\begin{table}[H]
\centering
\resizebox{\columnwidth}{!}{
\renewcommand{\arraystretch}{1.1}
\begin{tabular} {l||c|c|c|c}
&\textbf{Individual}&\textbf{Processed}&\textbf{Aggregated}&\textbf{Global}\\
\hline
Location&[1, 3, 9]&[1, 2, 7]&[1, 3, 10]&[1, 3, 8]\\
Communications&[.95, 2, 5.96]&[.9, 2, 8]&[1, 2, 8]&[1, 2, 7]\\
Apps &[1, 2, 6]&[1, 2, 5]&[1, 1, 5]&[1, 2, 5]\\
Media &[.5, 1, 5]&[.5, 1, 3]&[.5, 1, 5]&[.5, 1, 4]\\
\end{tabular}
}
\caption{[Q1,median,Q3] triplets for bid values (\euro{}) per category and level of complexity.}
\label{tab:bid-values}
\end{table}

\subsection{Awards}
The total amount won by participants in the form of auction awards was \euro{262} which was paid in Amazon vouchers. 

Additionally, we selected the ten subjects with the highest 
response rate and ran a raffle to select the winner of a final prize of \euro{100}.

A total of 29 subjects won at least one auction during the study; the cardinality of the winning set ramped from 5 to 29 as an effect of the increased number of auctions run in 
the 
second phase of the study. 

\subsection{Participation}
The participation rates for daily surveys is 79\%.

As mentioned earlier, users were granted opt-out options for each survey question by ticking a check-box which portrayed \texttt{"I do not want to sell this information"}. 

Table~\ref{tab:rr-stats} reports statistics of opt-out and distributions of valid responses (\emph{i.e.} survey items for which participants did not opt-out and entered their bid) 
for each category.
 
\begin{table*}[!ht]
\centering
\resizebox{\linewidth}{!}{
 \begin{tabular}{l||c|c|c|c|c||c|c}
  \textbf{Category}&\textbf{1st quartile}&\textbf{median}&\textbf{3rd quartile}&\textbf{mean}&\textbf{st\_dev}&\textbf{Opt-out (avg.\%)}&\textbf{Opt-out (median\%)}\\
\hline
  Location&34.3&69.8&84.3&58.2&33.1&17.7&2.63\\
  Communications&55.8&74.4&88.4&64.8&29.9&5.01&0\\
  Running Apps&40.7&65.1&81.4&58&30.8&7.59&0\\
  Media&62.8&76.7&90.7&66.4&32.8&9.25&0\\
 \end{tabular}
}
\caption{Distribution statistics of valid bid responses per category. Values reported in percentages. Last columns portray the opt-out statistics per category.}
\label{tab:rr-stats}
\end{table*}

\section{Data Analysis}
\label{sec:analyses}
The bidding data that was collected in the study is not normally distributed. Hence, we applied non-parametric analysis to test whether significant differences exist in the value 
distributions 
of different types of personal data. Thus, we report results using the Kruskal-Wallis test with a level of significance of $p<.05$.

Furthermore, we carried out correlation analyses to investigate whether associations between mobile phone usage patterns, demographics, subjects' predispositions, traits and 
auction 
behavior exist. For these analyses we employed the non-parametric Spearman's Rho method with a level of significance of $p<.05$.

\subsection{Bids}
We investigate first daily bids and specifically 
whether significant differences exist between (1) the categories and (2) levels of complexity within each category of mobile personal data we collected.

\subsubsection{Between-Category Study}
Significant differences in bid distributions were found between all data categories, with the only exception of Communications and Apps. 

The lack of statistically significant differences between Communications and Apps could be partially 
explained by the fact that most of the apps installed and used  
by participants in the study are communication apps. In terms of both running time and installations, $\sim$50\% of 
the top 20 apps are messaging apps (WhatsApp and similar), email (Gmail, Hotmail, Y!Mail), voice-over-IP clients (Skype, Viber) and social networking clients (Facebook).
We thus hypothesize that the distinction between Communication and Apps might be blurred.
We leave the validation of such a hypothesis to future work.
Nonetheless, the finding that participants seem to perceive, and consequently valuate, communications provided by a telco company and those provided by mobile apps in a similar 
manner, is intriguing and worth investigating.

\subsubsection{Within-Category Study}
Next we analyzed the differences in the distribution of bids within the different levels of complexity of mobile personal data. In other words, we looked if bid distributions 
within a given mobile data category showed significant differences for individual, aggregated, and processed information.

\paragraph{Applications.}
Significant differences emerged between individual and aggregated information ($p=.0108$), and between aggregated and processed 
information ($p=.039$). In particular, aggregated information about running applications (e.g. \emph{yesterday 7 applications were running on your device}) was valued less 
($\tilde x=$ \euro{1}) than individual (e.g. \emph{yesterday the Gmail application was running on your device}) or processed (e.g. \emph{yesterday the Gmail application ran
for 120 minutes on your device}) information ($\tilde x=$ \euro{2}). 
No significant difference was found between monetary valuations of individual and processed information on running applications ($p=0.659$).

\paragraph{Media.} Within the Media category, a significant difference in bid distributions was found ($p=.046$) between aggregated (e.g. \emph{yesterday you shot 8 pictures}) and 
processed (e.g. \emph{yesterday night you shot 3 pictures}) information. While for both information types the median bid value is $\tilde x=$ \euro{1}, a significant difference 
exists in terms of dispersion: the quartile coefficient of dispersion (\emph{i.e.} the ratio between difference and sum of the 3rd and 1st quartiles) is, respectively, 
$qcod_{agg}=.81$ and $qcod_{pro}=.71$.

\paragraph{Communications and Locations.} No significant differences were found in within-category analyses for Communications and Locations. In other words, participants valued 
similarly the communication and location data with each of the 3 levels of complexity.

\subsection{Impact of the Change in Auction Strategy}
As described earlier, in the middle of the study we increased the frequency of auctions from 
weekly (phase 1) to daily (phase 2). 
This change was transparent to participants and the frequency of email 
interventions was kept constant -- every Thursday.
We designed these two phases to assess if the probability of winning
had an effect on bidding behavior.

Indeed, we observe significant differences in bid distributions between the two phases for all categories: locations ($p=.02$), communications ($p=.01$), apps ($p=.001$) and 
media 
($p=.005$). 
Moreover, we find that mobile PII valuations drop for all categories in the second phase, as more participants won the auctions to monetize their data.

\subsection{The Value of Bulk Information}
The monetary valuations gathered in the final questionnaire for bulk information (\emph{i.e.} all the data collected in the 6-weeks presented in aggregated/anonymized form) are 
summarized in Table~\ref{tab:bulk-bids-stats}. 
Since participants could opt-out, we also report opt-out percentages for bulk information. 
\begin{table}[H]
\centering
\resizebox{\columnwidth}{!}{
\begin{tabular}{c||c|c|c|c}
 &\textbf{Location}&\textbf{Communications}&\textbf{Apps}&\textbf{Media}  \\
 \hline
mean&588.1&51.1&170.4&25.1\\
median&22.5&15&20&5\\
opt-out (\%)&16.67&3.34&0&8.34\\
\end{tabular}
}
\caption{Median/mean values (\euro{}) for bulk bids, and corresponding opt-out percentages.}
\label{tab:bulk-bids-stats}
\end{table}

Comparing with daily bids (see Table \ref{tab:bid-values}), the median bids for bulk 
information are one order of magnitude larger than the median individual bids, except for the media category. Mean opt-out percentages are similar except for the apps category.
The value ranking obtained from daily bids 
(Location $>$ Communications $>$ Apps $>$ Media) is different from 
that obtained in bulk bids 
(Location $>$ Apps $>$ Communications $>$ Media). 
In particular, application-related bulk data 
is valued significantly higher than communications-related bulk information.

\subsection{Relationship between Bids and Daily Behaviors}
In order to assess whether significant effects exist between mobile phone usage patterns and bidding behavior, we first computed daily behavioral variables from the sensed data. 
Table~\ref{tab:usage-feats} depicts the variables that we extracted with a daily granularity and for each participant.
\begin{table}[H]
\centering

\resizebox{\columnwidth}{!}{
 \begin{tabular}{l||r}
  \textbf{Category}&\textbf{Daily Behavioral Variables}\\
\hline
  \multirow{4}{*}{Location}&Distance \emph{total/mean/median/std}\\
  &Speed \emph{mean/median/std}\\
  &Radius of Gyration\\
  &Number of Places Visited\\
\hline
  \multirow{3}{*}{Communications}&Calls Duration \emph{total/mean/median/std}\\
  &Calls Diversity\\
  &Calls Total\\
\hline
  \multirow{2}{*}{Applications}&Total Apps Running\\
  &Total Apps Running Time\\
\hline
  \multirow{1}{*}{Media}&Total Pics shot\\
 \end{tabular}
}
\caption{Daily behavioral variables computed from mobile phone usage data.}
\label{tab:usage-feats}
\end{table}

With respect to \emph{location} data, information about the number of places visited was derived under the assumption that two locations would correspond to different places if 
the 
distance between them was larger than a threshold set to 100 meters. The radius of gyration corresponds to the radius of the smallest circle encompassing all location points 
registered each day.

For all these behavioral variables, we computed higher-order features corresponding to their statistical behavior over the 6-weeks period: mean, median, standard deviation, 
coefficient of variation (ratio of the standard deviation to the mean) and the quartile coefficient of dispersion. The last two features capture dispersion effects.

Furthermore, for each participant and data category, we computed mean, median, and standard deviation of their bids.

\subsubsection{Daily Bids}
We studied all correlations found between daily behavioral variables and bids in each category. 

We found a positive correlation between the mean location bid value and the median of daily distance traveled ($R=.294, p=.024$). That is, the larger the daily distance traveled, 
the higher the valuations of location information. With respect to applications, there are several statistically significant correlations. 
In particular, the total app running time is negatively correlated with the median app bid value ($R=-.26, p=.048$), meaning that the more time a participant spent using mobile 
apps, the lower the median valuations of app information. No significant correlation was found between communication and photo-taking behavioral features and bids on the 
communications and media categories.

\subsubsection{Bulk Bids}
There were a number of significant correlations between bids on bulk information and daily behaviors. Below we summarize the most notable correlations that we found.

Mobility information was positively correlated with bids on bulk \emph{location}, \emph{communication} and \emph{application} information. In particular, with the median of the i) 
radius of gyration ($R=.46, p=.0008$ for loc.; $R=.37, p=.005$ for comm.; $R=.34, p=.009$ for apps); and ii) daily mean speed ($R=.29, p=.04$ for loc.; $R=.39, p=.002$ for comm.; 
$R=.29, p=.029$ for apps). Location and application data was also positively correlated with the median of the daily mean distance traveled ($R=.39, p=.005$ for loc.; $R=.28, 
p=.031$ for apps) whereas communication bids were also positively correlated with the median of the i) total distance traveled ($R=.314, p=.018$) and ii) number of places visited 
($R=.336, p=.011$).

We also found statistically significant negative correlations of bulk location, communication and application bids with the coefficients of variation of mobility variables.

These correlations imply that the larger the daily distance traveled, the higher the valuation of  location, communication and application bulk bids. Conversely, the higher the 
variation in the patterns of mobility of a person, the lower his/her valuation of location, communication and app bulk information.
Note that bulk communication bids were not correlated with communication variables. 

In addition, bulk application bids are negatively correlated with the cumulative sum of daily unique total apps ($R=-.37, p=.003$) and with the median 
($R=-.28, p=.029$) and mean ($R=-.26, p=.04$) of total apps running daily. 

Finally, bulk media bids are correlated with the cumulative sum of 
daily unique total apps ($R=-.29, p=.03$).

\subsection{Relationship between Bids, Demographics, Traits and Dispositions}
\subsubsection{Daily Bids}
In the case of daily bids, we did not find any meaningful statistically significant correlation between bids and our participants' demographics or personality. 

There were statistically significant correlations with \emph{self-disclosure} variables that could be explained by the relevance of privacy aspects for all types of 
self-disclosure 
\cite{MeschBeker2010}.
In particular, the Intentional/Unintentional factor in self-disclosure is positively correlated with bids in three categories (communication, applications and media): (1) mean 
($R=.258, p=.048$), median ($R=.291, p=.02$) and standard deviation  ($R=.323, p=.012$) in communication bids, (2) median application bid value ($R=.26, p=.04$), and (3) median 
($R=.30, p=.02$), mean ($R=.27, p=.041$), and standard deviation ($R=.305, p=.019$) of media bids. 

\subsubsection{Bulk Bids}
Bulk location bids are found to be negatively correlated with Creativity ($R=-.375, p=.007$), while having positive correlations with the Intentional/Unintentional 
factor in self-disclosure ($R=.295, p=.039$) and Agreeableness ($R=.31, p=.027$). 
Interestingly, a positive correlation exists between bulk location bids and personal income 
($R=.32, p=.02$).
Furthermore, bulk communication information positively correlates with Agreeableness ($R=.31, p=.018$), and with the Intentional/Unintentional factor in self-disclosure ($R=.34, 
p=.009$).

\section{Insights from the EoS survey}
In the final survey, we asked our participants about particular bids they made during the 6-week data collection phase, and gave them the opportunity 
to express their views and concerns in free-form text (see Table~\ref{tab:finalsurvey} for details).

\subsection{Trust}
As seen in Table \ref{tab:finalsurvey}, Q4 asked our participants about 
their trust preferences with respect to 5 different entities who could be the safekeepers 
of their personal data: themselves, banks, telcos, governments and insurance companies.
From the trust rankings provided by our participants, 
we computed a \emph{trust score} for each entity by assigning 
a 1 to 5 value according to its rank and subsequently normalizing by the number 
of respondents. The final ranking that we obtained was: \emph{yourself} (.997), \emph{banks} (.537), \emph{telcos} (.513), \emph{government} (.49), and \emph{insurance companies} 
(.46). 

This result is aligned with 
the initial survey answers (Q2 and Q3 in Table~\ref{tab:initsurvey}) 
where participants conveyed that they do not trust telco operators or app providers with how they use their data.

In sum, overwhelmingly our participants trust themselves with their personal data more than any other entity, followed by banks and telcos. Insurance companies were the least 
trusted party. A similar question was also asked by Carrascal \emph{et al.} \cite{carrascal2013} obtaining similar results: the most trusted entity for a subject was the subject 
himself and the least trusted entities were the insurance companies. Interestingly, in our study, conducted in Italy, government was the second \emph{least} trusted entity while 
in 
Carrascal \emph{et al.} \cite{carrascal2013}, conducted in Spain, the government was the second \emph{most} trusted entity. 

\subsection{Lowest/Highest Bids per Category}
When analyzing the lowest/highest bids per category, we found that ~70\% of the highest bids for all categories took place in the first phase of the study (during the first three 
weeks). Adding more auctions (as it happened in the second phase of the study) led to lower bids. 

In the communications category, 61\% of the time participants entered a low bid to win and sell the associated communications information. This was significantly higher than for 
any other category. For all other categories, the most common reason reported for entering the low bid was that the information was not important. This finding suggests that 
participants found communication data to be the most desirable to sell.

Conversely, location was the most sensitive category of information as 25\% of the time participants entered a high location bid in order to avoid selling the information. This 
was 
significantly higher than for the other categories (5\% for communications, 3\% for apps and 6\% for media).  

\subsection{Insights about Opt-out Choices}
Location was the category of data for which subjects opted-out the most (56\%), followed by media (24\%), apps (18\%) and communications (2\%).
In the free-text explanations provided by our subjects it is clear that location is deemed
to be the most sensitive category of information, \emph{e.g.}: 
\begin{quote}
 \emph{``I don't like the idea of being geo-localized.''}\\
 \emph{``This kind of information is too detailed and too personal.''}
\end{quote}

Interesting explanations were also provided to justify the choice of not selling apps information, including that from apps usage is possible to infer information related to 
interests, opinions (expecially political opinions), and tastes:
\begin{quote}
 \emph{``From the usage of some applications it is possible infer information such as political orientation and other opinions and interests.''}
\end{quote}

\section{Discussion and Implications}
From the previously described analyses we can draw six insights related to mobile personal data:

\subsection{The Value of Bulk Mobile PII}
Carrascal \emph{et al.} \cite{carrascal2013} have reported higher values in their study on valuation of personal Web-browsing information than the ones we obtained in our study. 
The overall median bid value in our study was $\tilde x=$ \euro{2} while Carrascal \emph{et al.} reported an overall median bid value equal to $\tilde x=$ \euro{7} when they took 
in account context-dependent personal information. 
There are a few methodological differences between both studies which might 
explain the differences in bid values. In particular, \cite{carrascal2013} 
asked participants to provide a valuation of personal information captured while 
browsing the Web \emph{in-situ} using a rSEM methodology. Instead, we 
employed a DRM methodology querying participants about ther mobile PII 
from the previous day. 
From the valuations obtained in \cite{carrascal2013} and our study, it seems that individual pieces of PII are not as valuable when queried 
\emph{out-of-context} --such as in our study-- than \emph{in context} --such as in \cite{carrascal2013}. 

Conversely, bulk mobile PII was valued higher in our study 
than in \cite{carrascal2013} and significantly higher than individual PII. 
As shown in Tables \ref{tab:bid-values} and \ref{tab:bulk-bids-stats}, bulk information was valued an order of magnitude higher than individual data except for information in the 
media category. 
This finding is probably due to the power of the visualizations in the EoS survey, 
particularly for location and apps data. 

One hypothesis for this higher valuation 
is that participants realized how bulk data conveyed information about their life-style and habits and therefore considered it to be more valuable than daily items. Recently, Tang 
\emph{et al.} have shown the impact of different visualization types (text-, map-, and time-based) on social sharing of location data \cite{TangEtAl2011}. 

This result has a direct consequence for the design of trading mobile PII and highlights an asymmetry between buyers and sellers: for buyers, it would be more profitable to 
implement mechanisms to trade single pieces of information --that they could later aggregate. For sellers, however, it would be more advantageous to sell bulks of information.

\subsection{Location, location, location} 
As shown in Tables \ref{tab:bid-values} and \ref{tab:bulk-bids-stats}, location information received 
the highest valuation for all levels of complexity and was the most opted-out 
category of mobile PII. Bulk location information was very highly valued, 
probably due to the powerful effect of the map visualization in the EoS survey.
Several participants also expressed that they did not 
want to be geolocalized and considered location information to be highly sensitive and personal. 

Moreover, we found statistically significant correlations between mobility behaviors (\emph{e.g.} mean daily distance traveled, daily radius of gyration, etc.) and valuations of 
personal data. Not all users value their personal data equally: the more someone travels on a daily basis, the more s/he values not only her/his location information but also 
her/his communication and application information. 

Regarding this relation, previous works who focused on location information have presented contrasting results
~\cite{CvrcekEtAl2006,DanezisEtAl2005}; as we probe participants daily 
about fine-grained personal data they have just produced, our approach substantially differs from these survey-based studies, and it is thus difficult to directly compare with 
these works. Generally, our results seem to support the findings presented in~\cite{DanezisEtAl2005}.

These insights may have an impact on the design of commercial 
location-sharing applications. While users of such applications might consent
at install time to share their location with the app, our work suggests that 
when explicitly asked about either individual or bulk location data, $\sim 17\%$ of users
decide not to share their location information. In addition, mobility behaviors will influence the valuations of PII. 

Tsai \emph{et al.} \cite{TsaiEtAl2010} conducted an online survey with more than 500 American subjects to evaluate their own perceptions of the likelihood of several 
location-sharing scenarios along with the magnitude of the benefit or harm of each scenario (\emph{e.g.} being stalked or finding people in an emergency). The majority of the 
participants found the risks of using location-sharing technologies to be higher than the benefits.
However, today a significant number of very popular mobile apps such as Foursquare and Facebook Places make use of location data. These popular commercial location sharing apps 
seem to mitigate users' privacy concerns by allowing them to selectively report their location using check-in funtionalities instead of tracking them automatically. 

Based on our findings and given our participants concerns and high valuations of bulk location information, we believe that further user-centric studies on sharing and monetary 
valuation of location data are needed. 

\subsection{Socio-demographic characteristics do not matter, behavior does} 
When we correlated bid values against socio-demographic characteristics, we did not find significant correlations. This result is in contrast to previous work that found 
socio-demographic (mainly sex and age) differences in privacy concerns and consequent information disclosure behaviors \cite{ChristofidesEtAl2011, FogelNehmad2009}.
However, these previous studies were focused mainly on online information and on disclosure attitudes and privacy concerns than on monetary valuation of personal data. Carrascal 
\emph{et al.} \cite{carrascal2013}, instead, found results in line with ours (no significant correlations) except for a surprising low valuation of online information from older 
users.

On the other hand, we found statistically significant correlations between behavior (particularly
mobility and app usage) and valuations of bids. From our findings it seems that personal 
differences in valuations of mobile PII are associated with behavioral differences 
rather than demographic differences. In particular, the larger the daily distance traveled
and radius of gyration, the higher the valuation of PII. 
Conversely, the more apps a person used, the lower the valuation of PII. 
A potential reason for this correlation is
due to the fact that savvy app users have accepted that mobile apps collect
their mobile PII in order to provide their service and hence value their mobile PII
less.

\subsection{Intentional self-disclosure leads to higher bids} 
We found a positive correlation between the Intentional/Unintentional dimension of self-disclosure and the median values of the bids. This result could be explained by the fact 
that people with more intentional control about disclosing their own personal information, may be more aware of their personal data and hence also value it more from a monetary 
point of view.

Interestingly, we did not find significant correlations between bid values and other traits with the exception of Agreeablenness (with bulk location and communication bids).
Previous studies on the influence played by individual traits (usually personality traits and LoC) on privacy dispositions and privacy-related behaviors have provided contrasting 
evidence: some of them found small correlations \cite{Lo2010, QuerciaEtAl2012}, while Schrammel \emph{et al.} found no correlations \cite{SchrammelEtAl2006}. 
Hence, our results require additional investigations in order to clarify which are, if any, the dispositions and individual characteristics to take in account when a buyer makes a 
monetary offer for personal data.

\subsection{Trust}
From our study and from Carrascal \emph{et al.} \cite{carrascal2013}, it clearly emerges that individuals mainly trust themselves to handle their own personal data. This result 
suggests the adoption of a decentralized and \emph{user-centric} architecture for personal data management. 

Recently, several research groups have started to design and build personal data repositories which enable people to control, collect, delete, share, and sell 
personal data \cite{yves2012trusted,MunEtAl2010}, and whose value to users is supported by our findings.

\subsection{Unusual days lead to higher bids}
During our study there were two unusual days: December 8th (Immaculate Conception Holiday) 
and November 11th (a day with extremely strong winds which caused multiple road blocks and accidents). As can be seen in Figure~\ref{fig:medianbids}, 
the median bids for all categories in these two days were significantly higher than 
for the rest of the days in the study. Perhaps not surprisingly, participants in our study value
their PII higher in days that are unusual when compared to typical days. 

This result suggests that not all PII even within the same category and level
of complexity is valued equally by our participants, which has a direct implication 
for personal data markets and for services that monetize mobile personal data.

\section{Conclusion}
\label{sec:conclusions}
We have investigated the monetary value that people assign to 
their PII as it is collected by their mobile phone. In particular, we have taken into 
account four categories of PII (location, communication, apps and media) with three levels
of complexity (individual, aggregated and processed). 
We have carried out a comprehensive 6-week long study in a living lab environment with 60 participants adopting a Day Reconstruction Method along with a reverse second price 
auction mechanism to collect honest monetary valuations.

We have found that location is the most valued category of PII and that bulk information is 
valued much higher than individual information (except for the media category). We
have identified individual differences in bidding behaviors which are not correlated
with socio-demographic traits, but are correlated with behavior (mobility and app usage) and 
intentional self-disclosure. 

Finally, we have found that participants trust themselves 
with their PII above banks, telcos and insurance companies and that unusual days 
are perceived as \emph{more valuable} than typical days.

\bibliographystyle{abbrv}

\begin{thebibliography}{10}

\bibitem{AcquistiGrossklags2005}
A.~Acquisti and J.~Grossklags.
\newblock Privacy and rationality in individual decision making.
\newblock {\em IEEE Security \& Privacy}, 2:24--30, 2005.

\bibitem{AcquistiEtAl2009}
A.~Acquisti, L.~John, and G.~Loewenstein.
\newblock What is privacy worth.
\newblock In {\em Twenty first workshop on information systems and economics
  (WISE)}, pages 14--15, 2009.

\bibitem{AdarHuberman2001}
E.~Adar and B.~Huberman.
\newblock A market for secrets.
\newblock {\em First Monday}, 6(8), 2001.

\bibitem{aharony2011social}
N.~Aharony, W.~Pan, C.~Ip, I.~Khayal, and A.~Pentland.
\newblock Social fmri: Investigating and shaping social mechanisms in the real
  world.
\newblock {\em Pervasive and Mobile Computing}, 7(6):643--659, 2011.

\bibitem{Amichai2010}
Y.~Amichai-Hamburger and G.~Vinitzky.
\newblock Social network use and personality.
\newblock {\em Computers in Human Behavior}, 26(6):1289--1295, 2010.

\bibitem{AperjisHuberman2012}
C.~Aperjis and B.~A. Huberman.
\newblock A market for unbiased private data: Paying individuals according to
  their privacy attitudes.
\newblock {\em arXiv preprint arXiv:1205.0030}, 2012.

\bibitem{BarkhuusEtAl2008}
L.~Barkhuus, B.~Brown, M.~Bell, S.~Sherwood, M.~Hall, and M.~Chalmers.
\newblock From awareness to repartee: sharing location within social groups.
\newblock In {\em Proceedings of the SIGCHI Conference on Human Factors in
  Computing Systems}, pages 497--506. ACM, 2008.

\bibitem{carrascal2013}
J.~P. Carrascal, C.~Riederer, V.~Erramilli, M.~Cherubini, and R.~de~Oliveira.
\newblock Your browsing behavior for a big mac: Economics of personal
  information online.
\newblock In {\em Proceedings of the 22nd international conference on World
  Wide Web}, pages 189--200, 2013.

\bibitem{cherubini2009}
M.~Cherubini and N.~Oliver.
\newblock A refined experience sampling method to capture mobile user
  experience.
\newblock {\em CoRR}, abs/0906.4125, 2009.

\bibitem{ChristofidesEtAl2011}
E.~Christofides, A.~Muise, and S.~Desmarais.
\newblock Hey mom, what’s on your facebook? comparing facebook disclosure and
  privacy in adolescents and adults.
\newblock {\em Social Psychological and Personality Science}, 3(1):48--54,
  2012.

\bibitem{ConsolvoEtAl2005}
S.~Consolvo, I.~E. Smith, T.~Matthews, A.~LaMarca, J.~Tabert, and P.~Powledge.
\newblock Location disclosure to social relations: why, when, \& what people
  want to share.
\newblock In {\em Proceedings of the SIGCHI Conference on Human Factors in
  Computing Systems}, pages 81--90. ACM, 2005.

\bibitem{CostaMcCrae2005}
P.~T. Costa and R.~R. McCrae.
\newblock The revised neo personality inventory (neo-pi-r).
\newblock {\em The SAGE handbook of personality theory and assessment},
  2:179--198, 2008.

\bibitem{Cozby1973}
P.~Cozby.
\newblock Self-disclosure: A literature review.
\newblock {\em Psychological Bulletin}, 79(2):73--91, 1973.

\bibitem{CvrcekEtAl2006}
D.~Cvrcek, M.~Kumpost, V.~Matyas, and G.~Danezis.
\newblock A study on the value of location privacy.
\newblock In {\em Proceedings of the 5th ACM workshop on Privacy in electronic
  society}, pages 109--118. ACM, 2006.

\bibitem{DanezisEtAl2005}
G.~Danezis, S.~Lewis, and R.~J. Anderson.
\newblock How much is location privacy worth?
\newblock In {\em WEIS}, volume~5. Citeseer, 2005.

\bibitem{yves2012trusted}
Y.-A. de~Montjoye, S.~S. Wang, A.~Pentland, D.~T.~T. Anh, A.~Datta, et~al.
\newblock On the trusted use of large-scale personal data.
\newblock {\em IEEE Data Eng. Bull.}, 35(4):5--8, 2012.

\bibitem{FarmaCortivonis2000}
T.~Farma and I.~Cortivonis.
\newblock Un questionario sul "locus of control": suo utilizzo nel contesto
  italiano (a questionnaire on the locus of control: its use in the italian
  context.
\newblock {\em Ricerca in Psicoterapia}, 2, 2000.

\bibitem{FogelNehmad2009}
J.~Fogel and E.~Nehmad.
\newblock Internet social network communities: Risk taking, trust, and privacy
  concerns.
\newblock {\em Computers in Human Behavior}, 25(1):153--160, 2009.

\bibitem{Goodman1961}
L.~Goodman.
\newblock Snowball sampling.
\newblock {\em Annals of Mathematical Statistics}, 32:148--170, 1961.

\bibitem{HannEtAl2007}
I.-H. Hann, K.-L. Hui, S.-Y.~T. Lee, and I.~P. Png.
\newblock Overcoming online information privacy concerns: An
  information-processing theory approach.
\newblock {\em Journal of Management Information Systems}, 24(2):13--42, 2007.

\bibitem{HongLanday2004}
J.~I. Hong and J.~A. Landay.
\newblock An architecture for privacy-sensitive ubiquitous computing.
\newblock In {\em Proceedings of the 2nd International Conference on Mobile
  Systems, Applications, and Services}, MobiSys '04, pages 177--189. ACM, 2004.

\bibitem{HoyMilne2010}
M.~G. Hoy and G.~Milne.
\newblock Gender differences in privacy-related measures for young adult
  facebook users.
\newblock {\em Journal of Interactive Advertising}, 10(2):28--45, 2010.

\bibitem{HubermanEtAl2005}
B.~A. Huberman, E.~Adar, and L.~R. Fine.
\newblock Valuating privacy.
\newblock {\em Security \& Privacy, IEEE}, 3(5):22--25, 2005.

\bibitem{HuiEtAl2007}
K.-L. Hui, H.~H. Teo, and S.-Y.~T. Lee.
\newblock The value of privacy assurance: An exploratory field experiment.
\newblock {\em MIS Quarterly}, 31(1):19--33, 2007.

\bibitem{JensenEtAl2005}
C.~Jensen, C.~Potts, and C.~Jensen.
\newblock Privacy practices of internet users: Self-reports versus observed
  behavior.
\newblock {\em International Journal of Human-Computer Studies},
  63(1-2):203--227, 2005.

\bibitem{JunglasEtAl2008}
I.~A. Junglas, N.~A. Johnson, and C.~Spitzm{\"u}ller.
\newblock Personality traits and concern for privacy: an empirical study in the
  context of location-based services.
\newblock {\em European Journal of Information Systems}, 17(4):387--402, 2008.

\bibitem{kahnemanEtAl2004}
D.~Kahneman, A.~Krueger, D.~Schkade, N.~Schwarz, and A.~Stone.
\newblock A survey method for characterizing daily life experience: The day
  reconstruction method.
\newblock {\em Science}, 306:1776--1780, 2004.

\bibitem{KnijnenburgEtAl2013}
B.~P. Knijnenburg, A.~Kobsa, and H.~Jin.
\newblock Dimensionality of information disclosure behavior.
\newblock {\em International Journal of Human-Computer Studies},
  71(12):1144--1162, 2013.

\bibitem{KorzaanEtAl2009}
M.~Korzaan, N.~Brooks, and T.~Greer.
\newblock Demystifying personality and privacy: An empirical investigation into
  antecedents of concerns for information privacy.
\newblock {\em Journal of Behavioral Studies in Business}, 1, 2009.

\bibitem{LaneEtAl2010}
N.~D. Lane, E.~Miluzzo, H.~Lu, D.~Peebles, T.~Choudhury, and A.~T. Campbell.
\newblock A survey of mobile phone sensing.
\newblock {\em IEEE Communications Magazine}, 48(9):140--150, 2010.

\bibitem{LinEtAL2010}
J.~Lin, G.~Xiang, J.~I. Hong, and N.~Sadeh.
\newblock Modeling people's place naming preferences in location sharing.
\newblock In {\em Proceedings of the 12th ACM International Conference on
  Ubiquitous Computing}, pages 75--84. ACM, 2010.

\bibitem{LindqvistEtAl2011}
J.~Lindqvist, J.~Cranshaw, J.~Wiese, J.~Hong, and J.~Zimmerman.
\newblock I'm the mayor of my house: examining why people use foursquare-a
  social-driven location sharing application.
\newblock In {\em Proceedings of the SIGCHI Conference on Human Factors in
  Computing Systems}, pages 2409--2418. ACM, 2011.

\bibitem{Lo2010}
J.~Lo.
\newblock Privacy concern, locus of control, and salience in a trust-risk model
  of information disclosure on social networking sites.
\newblock In {\em AMCIS}, page 110, 2010.

\bibitem{MadanEtAl2012}
A.~Madan, M.~Cebri{\'a}n, S.~T. Moturu, K.~Farrahi, and A.~Pentland.
\newblock Sensing the "health state" of a community.
\newblock {\em IEEE Pervasive Computing}, 11(4):36--45, 2012.

\bibitem{MayerDavis1999}
R.~Mayer and J.~Davis.
\newblock The effect of the performance appraisal system on trust for
  management: a field quasi-experiment.
\newblock {\em Journal of Applied Psychology}, 84:123--136, 1999.

\bibitem{McAfeeMcMillan1987}
R.~Mc~Afee and J.~Mc~Millan.
\newblock Auctions and bidding.
\newblock {\em Journal of Economic Literature}, 25:699--738, 1987.

\bibitem{MeschBeker2010}
G.~Mesch and G.~Beker.
\newblock Are norms of disclosure of online and offline personal information
  associated with the disclosure of personal information online?
\newblock {\em Human Communication Research}, 36:570--592, 2010.

\bibitem{MunEtAl2010}
M.~Mun, S.~Hao, N.~Mishra, K.~Shilton, J.~Burke, D.~Estrin, M.~Hansen, and
  R.~Govindan.
\newblock Personal data vaults: A locus of control for personal data streams.
\newblock In {\em Proceedings of the 6th International Conference}, Co-NEXT
  '10, pages 1--12, 2010.

\bibitem{Pentland2012}
A.~Pentland.
\newblock Society's nervous system: Building effective government, energy, and
  public health systems.
\newblock {\em IEEE Computer}, 45(1):31--38, 2012.

\bibitem{PeruginiDiBlas2002}
M.~Perugini and L.~Di~Blas.
\newblock The big five marker scales (bfms) and the italina ab5c taxonomy:
  Analyses from an emic-etic perspective.
\newblock In B.~de~Raad~B. and M.~Perugini, editors, {\em Big Five Assessment}.
  Gottingen: Hogrefe and Huber Publishers, 2002.

\bibitem{QuerciaEtAl2012}
D.~Quercia, R.~Lambiotte, D.~Stillwell, M.~Kosinski, and J.~Crowcroft.
\newblock The personality of popular facebook users.
\newblock In {\em Proceedings of the ACM 2012 conference on computer supported
  cooperative work}, pages 955--964. ACM, 2012.

\bibitem{Riederer2011}
C.~Riederer, V.~Erramilli, A.~Chaintreau, B.~Krishnamurthy, and P.~Rodriguez.
\newblock For sale : Your data: By : You.
\newblock In {\em Proceedings of the 10th ACM Workshop on Hot Topics in
  Networks}, HotNets-X, pages 13:1--13:6, New York, NY, USA, 2011. ACM.

\bibitem{Rotter1965}
J.~Rotter.
\newblock Generalized expectancies for internal versus external control of
  reinforcement.
\newblock {\em Psychological Monographs}, 1965.

\bibitem{Rotter1967}
J.~B. Rotter.
\newblock A new scale for the measurement of interpersonal trust.
\newblock {\em Journal of personality}, 35(4):651--665, 1967.

\bibitem{SchrammelEtAl2006}
J.~Schrammel, C.~K{\"o}ffel, and M.~Tscheligi.
\newblock Personality traits, usage patterns and information disclosure in
  online communities.
\newblock In {\em Proceedings of the 23rd British HCI Group Annual Conference
  on People and Computers: Celebrating People and Technology}, pages 169--174.
  British Computer Society, 2009.

\bibitem{TangEtAl2011}
K.~P. Tang, J.~I. Hong, and D.~P. Siewiorek.
\newblock Understanding how visual representations of location feeds affect
  end-user privacy concerns.
\newblock In {\em Proceedings of the 13th international conference on
  Ubiquitous computing}, pages 207--216. ACM, 2011.

\bibitem{TochEtAl2013}
E.~Toch and I.~Levi.
\newblock Locality and privacy in people-nearby applications.
\newblock In {\em Proceedings of the 2013 ACM international joint conference on
  Pervasive and ubiquitous computing}, pages 539--548. ACM, 2013.

\bibitem{TsaiEtAl2010}
J.~Tsai, P.~Kelley, L.~Cranor, and N.~Sadeh.
\newblock Location sharing technologies: Privacy risks and controls.
\newblock {\em I/S: A Journal of Law and Policy for the Information Society},
  6(2):119--151, 2010.

\bibitem{TsaiEtAl2011}
J.~Y. Tsai, S.~Egelman, L.~Cranor, and A.~Acquisti.
\newblock The effect of online privacy information on purchasing behavior: An
  experimental study.
\newblock {\em Information Systems Research}, 22(2):254--268, 2011.

\bibitem{VanKleekEtAl2012}
M.~Van~Kleek, D.~A. Smith, N.~Shadbolt, and m.~schraefel.
\newblock A decentralized architecture for consolidating personal information
  ecosystems: The webbox.
\newblock In {\em PIM}, pages 177--189, 2012.

\bibitem{VarianEtAl2005}
H.~Varian, F.~Wallenberg, and G.~Woroch.
\newblock The demographics of the do-not-call list.
\newblock {\em IEEE Security \& Privacy}, 3:34--39, 2005.

\bibitem{WantEtAl2002}
R.~Want, T.~Pering, G.~Danneels, M.~Kumar, M.~Sundar, and J.~Light.
\newblock The personal server: Changing the way we think about ubiquitous
  computing.
\newblock In {\em In Proceedings of 4th International Conference on Ubiquitous
  Computing}, pages 194--209, 2002.

\bibitem{WhelessGrotz1976}
L.~Wheeless and J.~Grotz.
\newblock Conceptualization and measurement of reported self-disclosure.
\newblock {\em Human Communication Research}, 2:338--346, 1976.

\bibitem{WieseEtAl2011}
J.~Wiese, P.~G. Kelley, L.~F. Cranor, L.~Dabbish, J.~I. Hong, and J.~Zimmerman.
\newblock Are you close with me? are you nearby?: investigating social groups,
  closeness, and willingness to share.
\newblock In {\em Proceedings of the 13th international conference on
  Ubiquitous computing}, pages 197--206. ACM, 2011.

\bibitem{YoungQuanHaase2009}
A.~L. Young and A.~Quan-Haase.
\newblock Information revelation and internet privacy concerns on social
  network sites: a case study of facebook.
\newblock In {\em Proceedings of the fourth international conference on
  Communities and technologies}, pages 265--274. ACM, 2009.

\end{thebibliography}


\end{multicols}

\end{document}